\documentclass[11pt,a4paper]{article}
\usepackage[final]{arr}
\usepackage{times}
\usepackage{latexsym}
\usepackage{soul}
\usepackage{booktabs}
\usepackage{amsmath,caption}
\usepackage{colortbl}
\usepackage{arydshln}
\usepackage{multirow}
\usepackage{multicol}

\usepackage{microtype}
\usepackage{graphicx}
\usepackage{multirow}

\newcommand{\ours}[0]{\textsc{ForTaP }}

\title{\textsc{ForTaP}: Using Formulas for Numerical-Reasoning-Aware Table Pretraining}

\author{\textbf{Zhoujun Cheng}\textsuperscript{1}{\thanks{~~The first two authors contribute equally.}}~~, \textbf{Haoyu Dong}\textsuperscript{2}\footnotemark[1]~~{\thanks{~~Corresponding authors.}}~~,  \textbf{Ran Jia}\textsuperscript{2}, \\
  \textbf{Pengfei Wu}\textsuperscript{3}, \textbf{Shi Han}\textsuperscript{2}, \textbf{Fan Cheng}\textsuperscript{1}\footnotemark[2]~~, \textbf{Dongmei Zhang}\textsuperscript{2}\\
  \textsuperscript{1}MoE Key Laboratory of Artificial Intelligence, AI Institute, \\Shanghai Jiao Tong University, Shanghai 200240, China\\
  \textsuperscript{2}Microsoft Research Asia, China
  \textsuperscript{3}Fudan University, China\\
  {\tt \{blankcheng,chengfan\}@sjtu.edu.cn},~~{\tt 17307130207@fudan.edu.cn}   \\ 
  {\tt \{hadong,raji,shihan,dongmeiz\}@microsoft.com} \\
 }
  
\date{}
\begin{document}
\maketitle

\begin{abstract}
Tables store rich numerical data, but numerical reasoning over tables is still a challenge. 
In this paper, we find that the spreadsheet formula, a commonly used language to perform computations on numerical values in spreadsheets, is valuable supervision for numerical reasoning in tables.
Considering large amounts of spreadsheets available on the web, we propose \ours, the first exploration to leverage spreadsheet formulas for table pretraining. Two novel self-supervised pretraining objectives are derived from formulas, numerical reference prediction (NRP) and numerical calculation prediction (NCP). While our proposed objectives are generic for encoders, to better capture spreadsheet table layouts and structures, we build \ours upon TUTA, the first transformer-based method for spreadsheet\&web table pretraining with tree attention. \ours outperforms state-of-the-art methods by large margins on three representative datasets of formula prediction, question answering, and cell type classification, showing the great potential of leveraging formulas for table pretraining. The code will be released at \url{https://github.com/microsoft/TUTA_table_understanding}.
\end{abstract}


\section{Introduction} \label{sec:introduction}

\begin{figure}[t!]
    \begin{center}
    \includegraphics[width=3.02in]{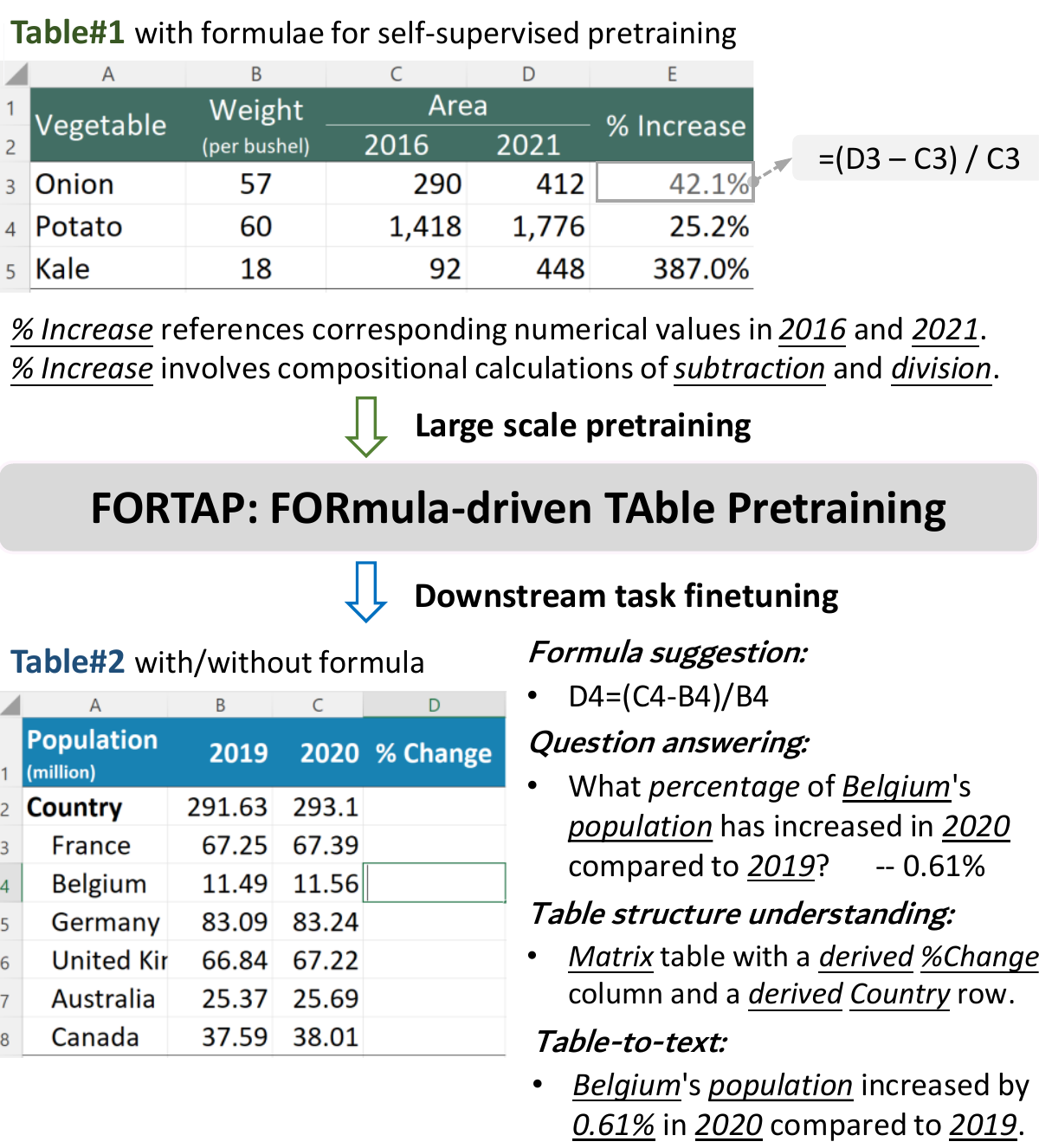}
    \end{center}
\vspace{-0.31cm}
\caption{It's desirable to learn numerical reasoning via formula pretraining and generalize it to various tasks.}
\label{fig:key idea}
\vspace{-0.4cm}
\end{figure}

Tables store rich numerical data, so a wide range of tasks require numerical reasoning over (semi-)structured tabular context, such as question answering over tables~\cite{chen2021finqa,zhu2021tat,cheng2021hitab}, table-to-text~\cite{suadaa2021towards,moosavi2021learning,cheng2021hitab}, spreadsheet formula prediction~\cite{spreadsheetcoder}, and table structure understanding~\cite{koci2019deco}. Take Table\#2 in Figure~\ref{fig:key idea} as an example, both suggesting the formula \texttt{(C4-B4)/B4} for cell D4 and answering ``0.61\%'' to the question require numerical reasoning capabilities of \textbf{(1)} understanding the contextual meaning of individual numerical cells, e.g., ``11.49'' at \texttt{B4} and ``11.56'' at \texttt{C4} are ``population"s of ``Belgium" in ``2019" and ``2020";
\textbf{(2)} inferring calculational relationships of numerical cells, e.g., percentage change from ``11.49'' to ``11.56''. As Figure~\ref{fig:key idea} shows, same capabilities also benefit table structure recognition and table-to-text. 
So it's a fundamental need to empower table modeling with stronger numerical reasoning capabilities.

However, it is challenging to endow a tabular model with robust numerical reasoning capabilities. First, understanding a local numerical cell needs dimension inference~\cite{chambers2008dimension}, unit inference~\cite{shbita2019parsing}, and index inference~\cite{dong2019semantic}, e.g., ``population'' (dimension), ``million'' (unit), ``2020'' (index), and ``Belgium'' (index) jointly describe ``11.56'' in Figure~\ref{fig:key idea}. It is non-trivial concerning the great flexibility of table semantic structures~\cite{tuta}.
Second, calculational relationships among two or more numerical cells are various and often compositional, e.g., ``F1 Score = 2 $\times$ (Recall $\times$ Precision) / (Recall + Precision)'' in machine learning papers and ``Profit Margin = Net Income / Sales'' in financial reports.
To make matters more challenging, \textbf{human labeling} for numerical reasoning in relevant tasks~\cite{chen2020open,suadaa2021towards,koci2019deco} is labor-intensive and error-prone, largely restricting the generalization ability of large models that are rather data-hungry.

Recently, table pretraining on large amount of unlabeled tables shows promising results on table understanding and reasoning. Self-supervised objectives are derived from tables and text such as Masked Language Models (MLM) ~\cite{tapas}, masked column prediction~\cite{tabert}, masked entity recovery~\cite{deng2020turl}, cell cloze and corrupt detection~\cite{tuta,tang2020rpt,iida2021tabbie}, table-text matching and alignment~\cite{wang2021retrieving,tuta,deng2020structure}.
However, numerical and calculational relationships of cells lack sufficient attention.
Then ~\cite{yoran2021turning} and ~\cite{tapex,yu2020grappa} synthesize questions and SQL queries, respectively, as training corpus for reasoning purpose, but SQL is only applicable to database-like relational tables, and importantly, it's challenging to ensure synthesized questions and SQLs be realistic, meaningful, and diverse.

Gladly, tens of millions of real spreadsheet formulas are publicly available on the web and can be valuable for numerical reasoning in tables. The spreadsheet formula is an expressive yet simple language consisting of operators (e.g., \texttt{+,/,\%}), functions (e.g., \texttt{SUM,MAX,COUNT}), referenced cells (e.g., \texttt{B4}), and constant values (e.g., $100$)~\cite{xlparser}. Since writing the formula \textbf{does not} require formal programming education, it's widely used by non-programmers such as business professionals or other kinds of domain specialists whose jobs involve computational tasks. 
So spreadsheet formulas cover real numerical calculations in a great variety of domains.

To this end, we propose FORmula-driven TAble Pretraining (\ours) for numerical reasoning. One should master two basic concepts to use the formula language: cells as variables and operators/functions as relationships between variables. So we explicitly decompose information in formulas into \textit{numerical reference} and \textit{numerical calculation} and devise two complementary tasks. Given a table as well as a formula cell in it, we mask the formula and then (1) 
the model classifies whether ``header $A$ references header $B$"
(we consider that ``header $A$ references header $B$" if the formula cell belonging to header $A$ references a numerical cell belonging to header $B$, as illustrated in Figure~\ref{fig:method}); (2) the model predicts the operator/function of two or more referenced numerical cells. 
Furthermore, to better encode and represent formulas, we also apply MLM to the token sequence of formulas. 

Considering the flexibility of table structures in spreadsheets, we base \ours on TUTA~\cite{tuta}, the first transformer-based method for spreadsheet tables with carefully-designed textual, numerical, positional, and formatting embedding layers. Importantly, its tree-based position encoding and attention are highly effective in representing generally structured tables. 
TUTA is pretrained with MLM, cell cloze, and table-text matching.

Experiment results on three tasks demonstrate that the significance of leveraging formulas for table pretraining. For formula prediction, \ours achieves $55.8\%$ top-1 accuracy, significantly surpassing TUTA ($48.5\%$), TaPEx~($43.2\%$), and SpreadsheetCoder ($40.4\%$) on Enron. For table question answering, TUTA achieves comparable accuracy with the best system on HiTab. After pretraining with formulas, \ours delivers a huge improvement of $+6.3\%$ as over previous SOTA, comparable to TaPEx. For cell type classification, on dataset DeEx, \ours largely improves TUTA by $+6.6\%$ on {\small \texttt{derived}} type and $+3.2\%$ on overall Macro-F1. 


\section{Preliminaries}

\subsection{TUTA as Encoder}
 TUTA~\cite{tuta} is the first pretraining architecture for spreadsheet tables. It is effective in capturing table semantic structures, achieving SOTA results on cell type and table type classification. As mentioned in Section~\ref{sec:introduction}, understanding table semantic structures is critical to numerical reasoning, so we choose TUTA to be the encoder of \ours. Since our pretraining tasks are generic for encoders of tables, future works can also explore other encoders such as ~\cite{tapas}. 
 
\vspace{3pt}
\noindent \textbf{Header Recognition.}
Headers usually provide short yet informative
descriptions of table contents in Natural Language (NL), so TUTA leverages the detected header regions and hierarchies, as presented in Section~\ref{sec:pretrain corpus}. ~\cite{spreadsheetcoder}
 also shows that using headers (even without considering hierarchies) greatly helps formula prediction. \ours follows to place detected headers in inputs.

\vspace{3pt}
\noindent \textbf{Architecture.} 
TUTA bases on BERT~\cite{devlin2019bert} with several enhancements: 
(1) a \textit{positional encoding layer} based on a unified \textit{bi-dimensional coordinate tree} to describe both the spatial
and hierarchical information of cells;
(2) a \textit{number encoding layer} to encode magnitude, precision, the first digit, and the last digit;
(3) a \textit{tree-based attention} mechanism that enables local cells to aggregate their structurally neighbouring contexts within a \textit{tree-based distance} threshold. 

\vspace{3pt}
\noindent \textbf{Model Input/Output.} 
The input consists of a table $T$ and optional NL texts $C$. By traversing the cell matrix of a table from left to right and from top to bottom, 
the input is linearized to ``$\texttt{\small[CLS]}$, $C_0$, ..., $C_{K-1}$, $\texttt{\small[SEP]}$, $T_{(0,0)}$, $\texttt{\small[SEP]}$, $T_{(0,1)}$, ..., $\texttt{\small[SEP]}$, $T_{(M-1,N-1)}$", where $K$ is the token length of NL texts, and $M$ and $N$ are the numbers of rows and columns of the table, respectively.
Note that $T_{(i,j)}$ refers to the token sequence of the cell string in the $(i+1)^{th}$ row and $(j+1)^{th}$ column, and each token has token, number, position, and format input embeddings.
The output of the encoder contains token-level, cell-level, and table-level embeddings. 
\ours follows these input/output settings except when inputting formula token sequence.


\subsection{Pretraining Corpus}  \label{sec:pretrain corpus}
\noindent \textbf{Spreadsheet Source and Preprocessing.}
We use the same spreadsheet table corpus as TUTA: (1) $13.5$ million public spreadsheet files are crawled from $1.75$ million websites;~(2) table ranges and headers are detected using TableSense~\cite{dong2019tablesense,dong2019semantic};~(3) header hierarchies are extracted with effective heuristics;~(4) extreme size tables are filtered out;~(5) duplicated tables are discarded. In the end, $4.5$ million spreadsheet tables are left. 

\vspace{3pt}
\noindent \textbf{Formula Preprocessing.}
Spreadsheet Formula is a widely-used end-user language for table organization and calculation. A formula consists of four types of formula tokens: operator~(e.g., \texttt{+,/,\%}), functions~(e.g., \texttt{SUM}), referenced cells~(e.g., \texttt{B4}) and constant values~(e.g., 100), which we denote as \texttt{OP}, \texttt{FUNC}, \texttt{CELL} and \texttt{CONST} in the rest part of the paper. We use XLParser~\cite{xlparser}, a highly-compatible formula parser with compact grammar, to analyze formula. In this way, we derive the AST of each formula~(an example AST in Figure~\ref{fig:method}) and the type of each formula token. Since we focus on single table setting, we discard the cross-table, cross-sheet, and cross-file formulas. Formulas with \textit{Array} or \textit{User-Defined-Function} are also discarded. The absolute reference sign ``\$" is deleted from formula strings, without changing their meanings. We only keep the first five occurrences of formulas in the same row/column because some spreadsheets contain hundreds of duplicated or dragged formulas in one row/column, which are inefficient for training. Formulas are linearized as formula token sequences in prefix 
representation of AST following SpreadsheetCoder~\cite{spreadsheetcoder}. 
Finally, 10.8 million formulas are derived.


\section{Pretraining Tasks} \label{sec:pretrain tasks}
As mentioned in Section~\ref{sec:introduction}, empowering table modeling with stronger numerical reasoning capabilities is a fundamental need.
Spreadsheet formulas naturally contain information of numerical references (\texttt{CELL}) and calculations (\texttt{OP}/\texttt{FUNC}), motivating us to devise effective tasks to leverage them for numerical-reasoning-aware pretraining.

\begin{figure}[t]
    \begin{center}
    \includegraphics[width=2.9in]{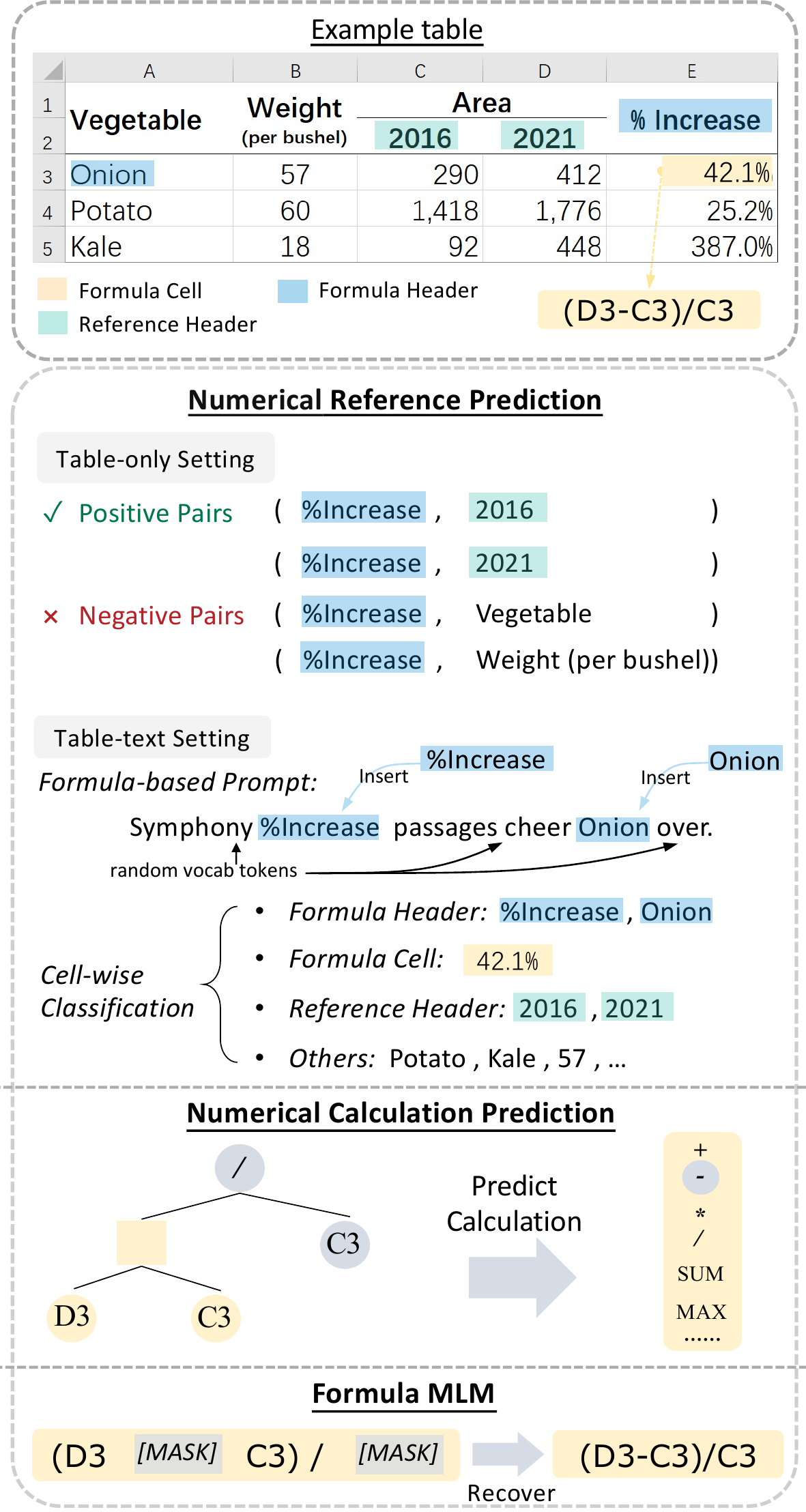}
    \end{center}
\vspace{-0.35cm}
\caption{An illustration of formula pretraining tasks.}
\label{fig:method}
\vspace{-0.38cm}
\end{figure}

Based on information parsed from the formula expression, we carefully devise two complementary objectives, Numerical Reference Prediction~(NRP) and Numerical Calculation Prediction~(NCP), to exploit the reasoning process behind referencing local cells (as operands) and applying calculations (on operands), respectively. 
Meanwhile, to get better representations of the spreadsheet formula, which could be further used in downstream applications like formula error detection~\cite{cheung2016custodes}, we extend MLM~\cite{devlin2019bert} from NL contexts to formulas. 
Figure~\ref{fig:method} gives an illustration of these tasks.

\vspace{3pt}
\noindent \textbf{Numerical Reference Predication (NRP)}
We consider ``header $A$ references header $B$" in a table if: in a formula, the formula cell~(cell with formula) belonging to header $A$ references a cell belonging to header $B$. Take the table in Figure~\ref{fig:method} as an example, the header \textit{``\%Increase"} references headers \textit{``2016"} and \textit{``2021"} since \texttt{E3} in column \textit{``\%Increase"} references \texttt{C3} and \texttt{D3} in columns \textit{``2016"} and \textit{``2021"}. We let the model learn header reference relationship since a cell belonging to a referenced header is more likely to be involved in the calculation. It is important but usually unknown as a priori, especially when tables are from diverse or unfamiliar domains. Note that we use header cells instead of data cells in this task since headers provide high-level descriptions of the data~\cite{spreadsheetcoder} and thus header reference relationships have more generic semantics across tables.

Given extracted header regions and hierarchies in corpus preprocessing, we first formulate NRP as a binary classification task over header pairs: given a formula cell $t_f$ and its referenced cells $\{t_p^{(i)}\}$, we first find their non-shared headers $h_f$ (for $t_f$) and $\{h_p^{(i)}\}$ (for $\{t_p^{(i)}\}$), then we group them as positive pairs $\{(h_f, h_p^{(i)})\}$. Usually a formula cell shares a header with referenced cells in the same row/column~(e.g., in Figure~\ref{fig:method}, \textit{``Onion"} is the shared header for \texttt{E3}, \texttt{C3}, \texttt{D3}). As it does not reflect header reference relationships, we exclude the shared header in this task. The negative pairs $\{(h_f, h_n^{(i)})\}$ are sampled among those unreferenced headers on the same direction~(either on top or left headers) of $h_f$~. Number of negative samples is at most 3:1 to positive ones to balance samples. The binary classification probability of the $i^{th}$ pair $p^{(i)}=f(\mathbf{h}_f, \mathbf{h}^{(i)}_{p/n})$, where $\mathbf{h}$ is the header cell embedding derived by the encoder and $f(\cdot)$ is a two-layer binary classification module. 

To inject table-text joint reasoning skills into \ours, which TUTA does not excel at, we further extend NRP task to table-text setting. Given a table with a formula cell, we first construct a formula-based prompt as context by picking $1$ to $10$ tokens randomly from the vocabulary as a noisy sentence and then inserting the row and column header of formula cell into it at random positions. Next, we jointly input the formula-based prompt and the table, and the task is to classify (1) formula header cell, (2) formula cell, (3) reference header cell, (4) other cells from the table. To precisely classify these cells, model needs to first align formula header cells in table with prompt~(alignment skill), then infer the intersection cell of formula header cells as formula cell~(spatial reasoning). Finally, it has to identify referenced cells~(numerical reasoning) by the formula headers.


The NRP loss $\mathcal{L}_{nr}$ is calculated as the sum of binary cross entropy loss and multi-class cross entropy loss under table-only and table-text setting.

\vspace{3pt}
\noindent \textbf{Numerical Calculation Prediction (NCP)}
 Given data cells as operands, a model then needs to find out which operators/functions should be applied. For example, in Figure~\ref{fig:method}, subtraction and division are applied on \texttt{C3} and \texttt{D3} in the formula. We hope the model can speculate the target operator/function based on the semantics, numeracy, and positions of given operands~(data cells). Thus, we design the task to predict the operator/function for a group of data cells with their contextual cell embeddings produced by the encoder.
 

We formulate it as a multi-class classification task: given a formula and its AST parsed in prerpocessing, we select the operators/functions $\{o^{(i)}\}$ satisfying that all direct children nodes $\{d^{(j)}\}^{(i)}$ on the formula AST of $o^{(i)}$ are in \texttt{CELL} type with integer or float data. The probability of predicting the operator/function of these data cells is $p^{(i)}=f(\texttt{POOL}(\{\mathbf{d}^{(j)}\}^{(i)}))$, where $\mathbf{d}$ is the output cell embedding by the encoder, $f(\cdot)$ is a two-layer classification module, and $\texttt{POOL}$ is a mean-pooling layer. Note that we only include the operator/function $o$ whose all direct children nodes are in \texttt{CELL} type in this task, because otherwise 
some descendant data cells will first be calculated via other operators/functions and thus have indirect connections with $o$~(e.g., in Figure~\ref{fig:method}, ``$/$" is not a target operator since its left child is an operator ``$-$"). We include $17$ common calculation operators/functions~(see Appendix~\ref{appendix:involved ops}) covered in spreadsheet formulas in this task. The NCP objective $\mathcal{L}_{nc}$ is the multi-class cross entropy loss.

\vspace{3pt}
\noindent \textbf{Formula MLM} \label{sec:formula mlm}
To encode formulas, we expand $41$ tokens in the vocabulary for all four formula token types, covering $99.1\%$ formulas in corpus. Added tokens are listed in Appendix~\ref{appendix:involved ops}. Note that a special case is the \texttt{CELL} type, like \texttt{D4}, because it references another cell.
Since referenced cells can be anywhere in a large table, it is infeasible to explicitly insert all cell positions into the vocabulary. Thus, for \texttt{CELL} type token in formula, we use a \texttt{[RANGE]} tag as input token and copy all cell-level embeddings (position, format, numeric, ...) from the referenced cell to this \texttt{CELL} type token.

We then apply MLM to formula tokens. Masking and recovering operators/functions is straightforward. When masking or recovering a referenced cell in a formula, we need to avoid label leakage from embeddings of the referenced cell. Thus, to mask a referenced cell, besides using the {\texttt{[MASK]}} token embedding, the number embedding is set to default to mask the number, and the position and format embeddings are set to the same as the formula cell. To recover a masked referenced cell $t_r$, the cell $t^{(i)}$ in input sequence with the highest probability $p^{(i)}=\texttt{Softmax}(f(\mathbf{t}_r, \mathbf{t}^{(i)}))$ is selected as the predicted cell, where $\mathbf{t}$ is output cell embedding of the encoder and $f(\cdot)$ is a two-layer classification module.
The objective $\mathcal{L}_{fmlm}$ is calculated as the sum of cross entropy loss over operator/function recovery and referenced cell recovery.

Finally, the total pretraining objective is 
\begin{align}
\mathcal{L} = \mathcal{L}_{nr} + \mathcal{L}_{nc} + \mathcal{L}_{fmlm}
\end{align}

\section{Experiments}
In this section, we describe the pretraining details and validate the effectiveness of \ours on three downstream tasks: formula prediction, question answering, and cell type classification. The statistics of datasets we use are listed in Table~\ref{tab:dataset stats}.

\subsection{Pretrain Implementation} \label{sec:pretrain impl}
We initialize \ours with parameters of the pretrained TUTA. 
The input is linearized following TUTA by concatenating the text~(the prompt built in NRP pretraining task) and the flattened table traversed in row order. Due to memory limit, we only place (1) header cells, (2) data cells on the same row/column of the formula cell, into the input sequence and skip the other cells. Our input pattern is reasonable as a tradeoff between performance and memory since we find that more than $89\%$ formulas only reference cells on the same row/column. To match different downstream tasks, for the cell with formula, we input its formula token sequence (e.g. \texttt{(C4-B4)/B4}) with $40\%$ probability, formula tag \texttt{[FORMULA]} with $30\%$~(the number embedding is set to default) and cell literal value with $30\%$ (e.g. number $42.1$). In experiments, we find it is more effective in Formula MLM to mask either all operators/functions or all referenced cells, so we implement it this way. We first pretrain $400K$ steps on sequence length $256$ with batch size $32$, and $250K$ steps on sequence length $512$ with batch size $8$. The whole pretraining phase takes about $4$ days on $4$ Tesla V100 GPUs. 


\begin{table}[t]
\begin{center}
\small
\scalebox{0.86}{
\begin{tabular}{l r r r}
\toprule[1.2pt]
\textbf{Dataset} & \textbf{Enron} & \textbf{HiTab} & \textbf{DeEx}\\
\midrule
\# samples~(train/dev/test) & $125k$  & $10.6k$ & $711k$ \\
~ & (formulas)  & (questions) & (cells)\\
\% hierarchical tables &  $51.0\%$ & $98.1\%$ & $43.7\%$\\
Avg. rows per table & $25.7$ & $17.1$  & $220.2$\\
Avg. columns per table & $12.4$ & $8.2$ & $12.7$\\
\midrule
Avg. formula sketch length  & $4.13$ & -\\

 Avg. op/func per formula & $1.62$ & -\\

\bottomrule[1.2pt]
\end{tabular}
}
\end{center}
\vspace{-0.35cm}
\caption{Statistics of downstream datasets.}
\vspace{-0.38cm}
\label{tab:dataset stats}
\end{table}

\subsection{Formula Prediction} \label{section:formula prediction}
Formula prediction~\cite{spreadsheetcoder} facilitates spreadsheet end-users by recommending formulas since writing formulas could be time-consuming and error-prone. Given a table and a target cell in table, the task is to predict a formula for the target cell. Formula prediction requires complex in-table numerical reasoning capabilities to predict both referenced cells and involved calculations. 

\vspace{3pt}
\noindent \textbf{Datasets.} Enron~\cite{hermans2015enron} is a massive database of public Excel Spreadsheet, containing over 17K spreadsheets with rich table structures and formula types. We exclude Enron from our pretraining corpus to prevent data leakage. Tables and formulas are preprocessed in the same way as the pretraining corpus. We divide Enron by sheet and the final dataset contains $100.3K$/$12.3K$/$12.9K$ table-formula pairs for train/dev/test. The formula cell in table is regarded as the target cell and the formula is seen as the ground truth in formula prediction task. We follow the evaluation metrics in SpreadsheetCoder~\cite{spreadsheetcoder}: (1) Formula Accuracy, (2) Sketch Accuracy, (3) Range Accuracy measuring the percentage of correctly predicted formulas, formula sketches~(formula using placeholder \texttt{[RANGE]} as referenced cells), and formula ranges~(only the referenced cells of formula).


Previous to our work, SpreadsheetCoder evaluates formula prediction on collected Google Sheets and Enron. However, we do not directly use its datasets for three reasons: (1) The Google Sheet corpus is not released, and for Enron,  SpreadsheetCoder only adopts formulas referencing cells within a limited rectangular neighborhood region ($21\times20$) of the formula cell, while we argue in real tables the referenced cells can be easily beyond this region. (2) A large proportion of table headers are not properly detected~(mentioned in its paper), while we adopt ranges and headers detected by TableSense~\cite{dong2019tablesense} and extract table header hierarchies. (3) Despite the inconsistencies above, we try to backtrack the original file to align with SpreadsheetCoder and apply our preprocessing. However, the document IDs of tables in SpreadsheetCoder are mostly empty. Thus, we build our dataset based on Enron and evaluate SpreadsheetCoder on it for a fair comparison.

\vspace{3pt}
\noindent \textbf{Baselines.} We adopt SpreadsheetCoder~\cite{spreadsheetcoder}, TaPEx~\cite{tapex}, and TUTA as our baselines. SpreadsheetCoder is a BERT-based model for formula prediction, incorporating headers and contextual information of neighbouring cells of the target cell. TaPEx is a BART-based~\cite{bart} table pretraining model, which implicitly learns a SQL executor.

\vspace{3pt}
\noindent \textbf{Fine-tune.} \ours consumes all header cells in the table and data cells lying on the same row/column of the target cell just like the manner in pretraining, with a max sequence length, $512$. The \texttt{[FORMULA]} tag is placed at the target cell position in input, whose number embedding is set to default. A two-stage LSTM formula decoder~\cite{dong-lapata-2018-coarse,spreadsheetcoder} accepts the formula cell embedding as input, and generates the formula by first generating formula sketches and then selecting referenced cells. All models in experiments are fine-tuned $800K$ steps on Enron. The beam size is $5$ for generating formula. Since SpreadsheetCoder only published part of its code, we re-implement it in PyTorch~\cite{paszke2019pytorch} based on its paper. Appendix~\ref{appendix:impl details} presents details about SpreadsheetCoder. TaPEx is built on BART model and thus naturally supports generation task. We follow the TaPEx table linearization strategy, assign the formula position in the source, and modify the target vocabulary as SpreadsheetCoder~\cite{spreadsheetcoder} to support generating referenced cells. We use the TaPEx-base model. It is fine-tuned for $30K$ steps~(converge at about $25K$) and evaluated on the checkpoint with the best dev performance.

\vspace{3pt}
\noindent \textbf{Results.} 
Table~\ref{tab:fp results} summarizes the results of formula prediction on the test set. As shown, \ours delivers a big improvement over SpreadsheetCoder by $+15.4\%$ and TaPEx by $+12.6\%$ on formula accuracy. 
We deduce that TaPEx falls behind TUTA and \ours because (1) the learnt executor may not be suitable for formula prediction, (2) it doesn't leverage hierarchical table structures. 
\ours also outperforms TUTA by $+7.3\%$, showing formula pretraining effectively assists formula prediction. 
 We also experiment under a low-resource setting~($20\%$ training data), and the improvements of \ours are more significant, surpassing TUTA by $+10.2\%$. 
Since Enron is not included in our pretraining corpus, this result well indicates formula pretraining can largely benefit formula prediction after seeing large numbers of real formulas. Moreover, we conjecture that formula pretraining potentially improves numerical reasoning capabilities of the model, because the two-stage prediction of formula sketches and ranges relies on numerical calculation and reference capabilities, respectively.

\begin{table}[t]
\begin{center}
\small
\begin{tabular}{l c c c}
\toprule[1.2pt]
(\%)  & \textbf{Formula} & \textbf{Sketch} & \textbf{Range} \\
\midrule
\specialrule{0em}{2pt}{0pt}
\multicolumn{4}{c}{\textit{20\% Train Set}} \\
\specialrule{0em}{0pt}{1pt} 
TUTA & $29.8$ & $50.5$ & $59.0$ \\
\midrule[0.2pt]
\ours & $\textbf{40.0}$ & $\textbf{57.6}$ & $\textbf{69.5}$ \\
\midrule[0.8pt]
\specialrule{0em}{2pt}{0pt}
\multicolumn{4}{c}{\textit{100\% Train Set}} \\
\specialrule{0em}{0pt}{1pt} 
SpreadsheetCoder & $40.4$ & $59.6$ & $67.7$ \\
TaPEx & $43.2$& - & - \\
TUTA & $48.5$ & $65.3$ & $75.3$ \\
\midrule[0.2pt]
\ours & $\textbf{55.8}$ & $\textbf{70.8}$ & $\textbf{78.8}$ \\
\bottomrule[1.2pt]
\end{tabular}
\end{center}
\vspace{-0.35cm}
\caption{Formula prediction accuracy on Enron.}
\vspace{-0.38cm}
\label{tab:fp results}
\end{table}

\subsection{Table Question Answering}
Table QA~\cite{pasupat2015compositional, cheng2021hitab} contains a table and an NL question over the table as the model input. Its output can be cell value(s) or number(s) calculated over numerical cell value(s). Table QA calls for both in-table numerical reasoning and table-text joint reasoning. 

\vspace{3pt}
\noindent \textbf{Datasets.} There are several datasets~\cite{pasupat2015compositional, cheng2021hitab, zhu2021tat, chen2021finqa} focusing on Table QA or Table-text hybrid QA. We choose to evaluate on HiTab~\cite{cheng2021hitab}, a hierarchical web table dataset for question answering and data-to-text. First, tables in HiTab contain rich table structures~($98.1\%$ tables are hierarchical) from $29$ domains, posing a challenge to numerical reasoning. Second, a large proportion of questions~($\sim40\%$) from Statistical Reports demands complex numerical inference over table and text. Moreover, questions in HiTab are revised from sentences written by professional analysts to ensure naturalness and meaningfulness. The QA evaluation metric is Execution Accuracy measuring the percentage of correctly predicted answers.

\vspace{3pt}
\noindent \textbf{Baselines.} We employ TaPas~\cite{tapas}, HiTab model~\cite{cheng2021hitab}, TaPEx~\cite{tapex}, and TUTA as our baselines. TaPas is an end-to-end table parsing model without generating logical forms, which enjoys pretraining on the large-scale table-text corpus from Wikipedia. HiTab devises a hierarchy-aware logical form for hierarchical tables, and predicts the answer using a weakly supervised semantic parser MAPO~\cite{mapo}, which is a reinforcement learning framework to systematically explore and generate programs. The question and table are encoded by BERT and the logical forms are generated by an LSTM decoder. TaPEx is introduced in Section~\ref{section:formula prediction}.

\vspace{3pt}
\noindent \textbf{Fine-tune.} 
We replace the BERT encoder of HiTab model with TUTA and \ours, and follow the fine-tuning settings of HiTab. We find that NRP pretrain task under table-text setting mentioned in Section~\ref{sec:pretrain tasks} is quite essential for QA performance and thus pretrain $80,000$ steps more with it on \ours in QA before fine-tuning. For TaPEx, we adopt the same table QA strategy in its paper by inputting the table and text as source, and generating the answer as target. The TaPEx-base model is trained for $20,000$ steps on HiTab.

\vspace{3pt}
\noindent \textbf{Results.} Table~\ref{tab:qa results} summarizes QA results on HiTab. \ours achieves SOTA ($47.0\%$) using MAPO as the semantic parser, surpassing the best system in HiTab paper with $+6.3\%$. Meanwhile, replacing BERT with TUTA does not see a significant performance gain. We conjecture one of the reasons is that TUTA may be not skilled at table-text joint reasoning, and \ours enhances this skill by the table-text setting of the NRP task. Finally, \ours performs comparatively with TaPEx, a recent pretraining tabular model as a powerful neural SQL executor targeting table reasoning. Note that this result is inspiring since \ours is pretrained on spreadsheet tables and can generalize to web table domain~(HiTab) with SOTA performance, indicating that the numerical reasoning skills learnt by \ours are robust to distinct scenarios.


\textbf{\begin{table}[t]
\begin{center}
\small
\begin{tabular}{l c c}
\toprule[1.2pt]
(\%)  & \textbf{Development} & \textbf{Test}\\
\midrule
TaPas & $39.7$ & $38.9$ \\
BERT~(\textit{MAPO}) & $43.5$ & $40.7$ \\
TUTA~(\textit{MAPO}) & $43.5$ & $41.3$ \\
TaPEx & $\textbf{48.8}$ & $45.6$ \\
\midrule[0.3pt]
\ours~(\textit{MAPO}) & $47.1$ & $\textbf{47.0}$ \\
\bottomrule[1.2pt]
\end{tabular}
\end{center}
\vspace{-0.2cm}
\caption{QA execution accuracy on HiTab. \textit{MAPO} means using MAPO+hierarchical-aware logical forms.}
\label{tab:qa results}
\end{table}}


\begin{table}[t]
\begin{center}
\small
\scalebox{0.82}{
\begin{tabular}{l c c c c c c c c}
\toprule[1.2pt]
(\%) & \textbf{M} & \textbf{N} & \textbf{Data} & \textbf{LA} & \textbf{TA} & \textbf{Derived} & \textbf{Avg.} \\
\midrule
\specialrule{0em}{2pt}{0pt}
$\text {CNN}^{\text {BERT}}$ & $76.3$ & $1.5$ & $95.2$ & $59.0$ & $75.4$ & $57.6$ & $60.8$\\
$\text {RNN}^{\text {C+S}}$ & $62.7$ & $40.8$ & $98.6$ & $56.9$ & $73.5$ & $48.8$ & $63.6$\\
TaBERT & $66.6$ &  $5.4$ & $94.3$ & $29.2$ & $59.2$ & $45.1$ & $50.0$\\
TaPas & $80.6$ & $20.3$ & $96.5$ & $56.9$ & $\textbf{90.1}$ & $56.6$ & $66.8$\\
TUTA & $\textbf{86.0}$ & $41.6$ & $99.1$ & $76.7$ & $82.0$ & $73.1$ & $76.4$ \\
\midrule[0.3pt]
\ours & $85.2$ & $\textbf{49.1}$ & $\textbf{99.3}$ & $\textbf{78.0}$ & $86.4$ & $\textbf{79.7}$ & $\textbf{79.6}$  \\
\specialrule{0em}{1pt}{0pt}
\specialrule{0em}{0pt}{1pt}
\bottomrule[1.2pt]
\end{tabular}
}
\end{center}
\caption{F1 scores of cell type classification on DeEx: \textbf{M}({\small \texttt{metadata}}), \textbf{N}({\small \texttt{notes}}), \textbf{Data}, \textbf{LA}({\small \texttt{left}} \textbf{}{\small \texttt{attribute}}), \textbf{TA}({\small \texttt{top attribute}}) , and \textbf{Derived}.}
\vspace{-0.2cm}
\label{tab:deex results}
\end{table}

\subsection{Cell Type Classification}
Cell type classification~(CTC)~\cite{koci2019deco,gol2019tabular, gonsior2020active} aims to interpret tabular data layouts automatically via classifying table cells by their roles in data layouts~(e.g., top attribute, data, derived). It requires understanding of table semantics, structures, and numerical relationships considering diverse table layouts.

\vspace{3pt}
\noindent \textbf{Datasets.} DeEx~\cite{koci2019deco} is a widely-studied CTC dataset with tables of various structures and semantics. DeEx includes tables from various domains by mixing three public corpora: Enron~\cite{hermans2015enron}, Euses~\cite{fisher2005euses}, and Fuse\cite{barik2015fuse}. Cells in DeEx are categorized into six fine-grained types: {\small \texttt{metadata, notes, data, left attribute, top attribute}}, and {\small {\texttt{derived}}}. The evaluation metric is the Macro-F1 score over all cell types.


\vspace{3pt}
\noindent \textbf{Baselines.} We compare \ours with two learning-based methods $\text {CNN}^{\text {BERT}}$\cite{dong2019semantic} and Bi-LSTM~\cite{gol2019tabular}, and three table-pretraining methods TaBERT~\cite{tabert}, TaPas~\cite{tapas}, and TUTA. 

\vspace{3pt}
\noindent \textbf{Fine-tune.} 
To handle large tables in DeEx, we split tables into chunks with a max input sequence length ($512$) and distribute headers to each chunk. 
For cells with formulas, \texttt{[FORMULA]} tags are used as input tokens. We fine-tune $100$ epochs on five folds and report the average scores. All these settings are the same as TUTA. 

\vspace{3pt}

Table~\ref{tab:deex results} lists the CTC results on DeEx. \ours achieves a SOTA Macro-F1 of $79.6\%$. Specifically, \ours largely improves the performance on type {\small \texttt{derived}} and {\small \texttt{notes}}, surpassing TUTA by $6.6\%$ and $7.5\%$. The improvement on {\small \texttt{derived}} indicates formula pretraining helps identifying cells derived by calculations over some other cells. Note that {\small \texttt{derived}} in DeEx not only includes cells with explicit formulas, but also those cells with hidden~(missing) formulas~\cite{koci2019deco}, which poses a great challenge to existing methods since it requires discovery of numerical relationships between cells. Thus, this is a strong signal that formula pretraining endows the model with better numerical reasoning capabilities. We think that the improvement on {\small \texttt{notes}} mainly benefits from the NRP pretraining task with formula-based prompts as the context, enhancing \ours's capability on table-text joint modeling.


\subsection{Analysis} \label{sec:analysis}
In this section, we analyze our method in terms of (1) the effects of different pretraining tasks, (2) whether and to what extent our model learns numerical reasoning skills.

\vspace{3pt}
\noindent \textbf{Effects of pretraining tasks.}
We conduct ablation studies on different pretraining tasks on the formula prediction task. Here we pretrain TUTA with each pretraining task and fine-tune on Enron dataset, as summarized in Table~\ref{tab:fp ablation}.  We can see that combining all pretraining tasks brings the most gain on formula accuracy. NRP and NCP improve more on range accuracy and sketch accuracy, respectively. This aligns with our design motivation that NRP targets on how to reference and NCP learns how to calculate. To our surprise, Formula MLM alone also largely benefits formula prediction. We deduce the reason is that both MLM and formula prediction requires encoding and recovering/generating capabilities of the formula token sequence.



\begin{table}[t]
\vspace{-0.3cm}
\begin{center}
\small
\begin{tabular}{l c c c}
\toprule[1.2pt]
(\%)  & \textbf{Formula} & \textbf{Sketch} & \textbf{Range}\\
\hline
\specialrule{0em}{2pt}{0pt}
TUTA & $48.5$ & $65.3$ & $75.3$ \\
TUTA\;+\;NRP  & $54.3$ & $69.0$ & $78.7$ \\
TUTA\;+\;NCP & $54.7$ & $\textbf{71.2}$ & $76.8$ \\
TUTA\;+\;FormulaMLM & $54.6$ & $70.2$ & $77.7$ \\
All~(\ours) & $\textbf{55.8}$ & $70.8$ & $\textbf{78.8}$ \\
\bottomrule[1.2pt]
\end{tabular}
\end{center}   
\vspace{-0.22cm}
\caption{Ablation study on formula prediction.}

\label{tab:fp ablation}
\end{table}

\vspace{3pt}
\noindent \textbf{Numerical reasoning skills.} We have shown our model learns numerical reasoning skills by two facts: (1) NRP and NCP improve more on the range and sketch accuracy on the formula prediction task, respectively; (2) our model boosts the performance of \texttt{\small {derived}} cell type on cell type classification. Here we further decompose QA accuracy of different operations on HiTab. The comparison between previous SOTA system BERT(MAPO) and our \ours(MAPO) is shown in Table~\ref{tab:qa operation acc}. As shown, our model improves most on complex cell selection~(cell indexed by $\geq3$ headers) and arithmetic~(e.g., \textit{difference}, \textit{sum}) problems. Note that complex cell selection not only requires table-text alignment, but also the references between headers considering that mentions of headers in question could be implicit or missing. Meanwhile, our model also handles superlative~(e.g., \textit{argmax}) and comparative~(e.g., \textit{less than}) problems better than BERT, despite these types are relatively infrequent in our formula pretraining corpus. To summarize, our model mainly improves numerical skills regarding cell reference and arithmetic, as well as other aspects like comparing and ranking.

\begin{table}[t!]
\vspace{-0.2cm}
\begin{center}
\small
\begin{tabular}{l l l }
\toprule[1.2pt]
\textbf{Operation}  & \textbf{BERT} & \textbf{\ours}\\
\hline
\specialrule{0em}{2pt}{0pt}
Complex Cell Selection & $48.4\%$ & $56.4\%$~($+8.0\%$) \\
Arithmetic & $6.0\%$ & $13.3\%$~($+7.3\%$) \\ 
Superlative & $22.7\%$ & $26.8\%$~($+4.1\%$) \\  
Comparative & $27.5\%$ & $30.5\%$~($+3.0\%$) \\
\bottomrule[1.2pt]
\end{tabular}
\end{center}   
\vspace{-0.35cm}
\caption{Accuracy on HiTab of different operations.}
\vspace{-0.38cm}
\label{tab:qa operation acc}
\end{table}

\section{Related Works}
\noindent \textbf{Table Pretraining.}
Table pretraining has been widely studied in recent years. Some works mine large-scale table-text pairs as pretraining corpus~\cite{deng2020turl, tabert, tapas, tuta}, some leverage annotated table-text datasets~\cite{strug, yu2020grappa}, and some synthesize a table-text corpus by templates~\cite{yu2020grappa, eisenschlos2020understanding}. Regarding pretraining tasks, they either train the model to recover masked tokens/column/cell/entity~\cite{tabert, tapas, tuta, deng2020turl}, or explicitly learn table-text alignments~\cite{strug, yu2020grappa}. Recently, TaPEx~\cite{tapex} adopts BART~\cite{bart} as a neural executor for synthesized SQLs to improve table reasoning. Whereas, our method explores to use real spreadsheet formulas to guide table pretraining.

\vspace{3pt}
\noindent \textbf{Numerical reasoning over Natural Language.} Numerical reasoning is important in NL domain~\cite{dua2019drop}. Numbers even account for $6.15\%$ of all unique tokens
in English Wikipedia~\cite{thawani-etal-2021-numeracy}. Various works target improving numerical reasoning skills on NL~\cite{andor2019giving, geva2020injecting, jin2021numgpt}. Except using pure NL, MathBERT~\cite{peng2021mathbert} pretrains NL documents with mathematical formulas. In this paper, we target numerical reasoning over (semi-) structured tables. 


\section{Conclusion}
In this paper, we present \ours, a numerical-reasoning-aware table pretraining model that learns numerical reasoning capabilities from spreadsheet formulas. Specifically, we design two pretraining tasks to capture numerical reasoning capabilities by explicitly predicting cell reference and calculation relations. Experiments show that \ours achieves new SOTA on formula prediction, question answering, and cell type classification. Further analyses indicate that formula pretraining indeed improves numerical reasoning skills of the model. One limitation of \ours is that we haven't fully exploit spreadsheet formulas beyond numerical reasoning. For example, logic functions like \texttt{VLOOKUP} and text functions like \texttt{LEN} can be leveraged to guide complex logic and text reasoning, which will be a promising direction in the future.


\section{Ethical Considerations}
In this work, we present a table pretraining method leveraging spreadsheet formulas.

\noindent \textbf{Dataset.} Our pretraing corpus is built upon public English spreadsheet files crawled from webs via the search engine~\cite{tuta}, covers various domains, and has been checked by a compliance team in a company to ensure that does not contain sensitive names or uniquely identifies individual people or offensive content. All datasets used for evaluation are licensed public datasets, e.g., for formula prediction, Enron~\cite{hermans2015enron} is a public spreadsheet dataset consisting of over $17$K spreadsheet files, and we re-purpose it for formula prediction following ~\cite{spreadsheetcoder}. 

\noindent \textbf{Application.} Our model shows its effectiveness in three representative table-related tasks. Formula prediction helps spreadsheet end-users to write formulas which could be tedious and error-prone. Table QA enables users to query on the table without the need of domain background knowledge. Cell type classification assists interpreting fine-grained table semantic structures, which help users to better understand table structures and contents. There may be risks that crooks use tabular models to automatically parse tables/forms to obtain private personal or company data in bulk, which should be prevented.

\bibliographystyle{acl_natbib}
\bibliography{acl_natbib}

\clearpage
\appendix

\section{Involved Operators/Functions of Formula} \label{appendix:involved ops}
We include $17$ common operators/functions in Numerical Calculation Prediction pretraining task, which consists of all the operators and four most commonly used aggregation functions in spreadsheet formula. The operators/functions are: $+$, $-$, $*$, $/$, $\wedge$, $\%$, $\&$, $=$, $\textless \textgreater$, $\textgreater$, $\textless$, $\geq$, $\leq$, \texttt{SUM}, \texttt{AVERAGE}, \texttt{MAX}, \texttt{MIN}.

To encode formula token sequence, we expand $41$ tokens in vocabulary for all four formula token types \texttt{OP, FUNC, CELL, CONST}, covering $99.1\%$ formulas in corpus. Here we list these tokens: (1) $1$ token for \texttt{CELL} token type: \texttt{[RANGE]}. (2) $3$ tokens for \texttt{CONST} token type: \texttt{[C-STR]}, \texttt{[C-NUM]}, \texttt{[C-BOOL]}. All constant tokens are categorized according to ``string", ``number", and ``bool". And they are replaced with these three tokens when encoding the formula. (3) $34$ tokens for \texttt{OP}/\texttt{FUNC} token type:  $[+]$($32.1\%$), \texttt{[SUM]}($20.6\%$), $[-]$($17.8\%$), $[/]$($6.7\%$), \texttt{[IF]}($2.6\%$), \texttt{[ROUND]}($1.2\%$), \texttt{[AVERAGE]}($1.2\%$), \texttt{[VLOOKUP]}($1.0\%$),  $[\textgreater]$($0.98\%$), $[=]$($0.79\%$), $[\textless]$($0.57\%$), \texttt{[ABS]}($\textless 0.5\%$), \texttt{[OFFSET]}, \texttt{[SUBTOTAL]}, \texttt{[MAX]}, $[\textless \textgreater]$, $[\wedge]$, \texttt{[LN]}, \texttt{[COUNTA]}, \texttt{[SQRT]}, \texttt{[MIN]}, \texttt{[ISERROR]}, \texttt{[EOMONTH]}, \texttt{[COUNT]}, \texttt{[AND]}, $[\%]$, \texttt{[INDEX]}, \texttt{[YEAR]}, \texttt{[MONTH]}, 
\texttt{[MATCH]}, $[\geq]$, \texttt{[MATCH]}, $[\leq]$, $[\&]$, \texttt{[UNKOP]}. The number in parentheses is the  ratio of \texttt{OP}/\texttt{FUNC} to the total number of \texttt{OP}/\texttt{FUNC} in corpus. Here \texttt{UNKOP} stands for unknown operator/function, similar to \texttt{[UNK]} in NL vocabulary. To distinguish formula \texttt{OP}/\texttt{FUNC} with some eponymous tokens in vocabulary~(e.g., ``sum", ``+"), we enclose formula \texttt{OP}/\texttt{FUNC} with square brackets. (4) special tokens \texttt{[START]}, \texttt{[END]}, $[:]$.

\section{Implementation Details} \label{appendix:impl details}
\noindent \textbf{More on Hyperparameters.} For pretraining, we first pretrain $400K$ steps with  max sequence length $256$, batch size $32$, then pretrain $250K$ steps with max sequence length $512$, batch size $8$. The whole pretraining phase is estimated to $3$ epochs, i.e., samples in the corpus are seen $3$ times in pretraining. The optimizer is Adam with learning rate $2e$-$5$. 

For formula prediction, we set max sequence length $512$ and fine-tune $800K$ steps with batch size $2$ on single GPU. The tokens beyond $512$ are truncated. If the formula cell is truncated~(rare case), we input the \texttt{[CLS]} embedding to the formula decoder. The two-stage decoder is first trained $100K$ for generating sketches, and then trained to generate sketches and ranges together. The optimizer is Adam with learning rate $2e$-$5$. 

For table question answering, we follow HiTab hyperparameters except that we find it is unnecessary to freeze encoder parameters at the first $5,000$ steps, so we train the encoder-decoder model together. 

For cell type classification, since some tables are extremely large in DeEx, we truncate the tables into sequences of max length $512$ by preserving the header cells~(both top and left) and traversing the data cells to fill the max sequence length. We fine-tune $100$ epochs on five folds with batch size $12$. The optimizer is Adam with learning rate $8e$-$6$.

\vspace{3pt}
\noindent \textbf{SpreasheetCoder} We implement SpreadsheetCoder mainly following its paper including the BERT-based table context~(row/column) encoder, two-stage decoder. One difference is that we did not implement the convolution layers for row and columns which is rather complicated . Instead, since SpreadsheetCoder uses convolution layer aiming to incorporate contextual information from different positions~(row/column), we explicitly add row embeddings and column embeddings~\cite{tapas} for input table tokens, which derives the similar accuracy gain of convolution layers~($4\%$ according to its paper), from $35.6\%$ to $40.4\%$ on Enron dataset. Furthermore, SpreadsheetCoder can only decode referenced cells in a rectangle window~($[-10, 10]$) of the target cell since it only keeps the formulas of this kind in dataset. We enable SpreadsheetCoder to predict referenced cells in a larger window which it can not solve by extending the vocabulary of range tokens from $[-10, 10]$ to $[-256, 256]$. Different from SpreadsheetCoder, \ours predicts ranges by selecting from input table cells instead of from a fixed cell vocabulary. In this way, theoretically~(without memory limit) our model can potentially predict referenced cells in an arbitrarily large table. Detailed error analysis of \ours on formula prediction is in Appendix~\ref{appendix:fp error analysis}.

\begin{figure}[t]
    \begin{center}
    \includegraphics[width=2.9in]{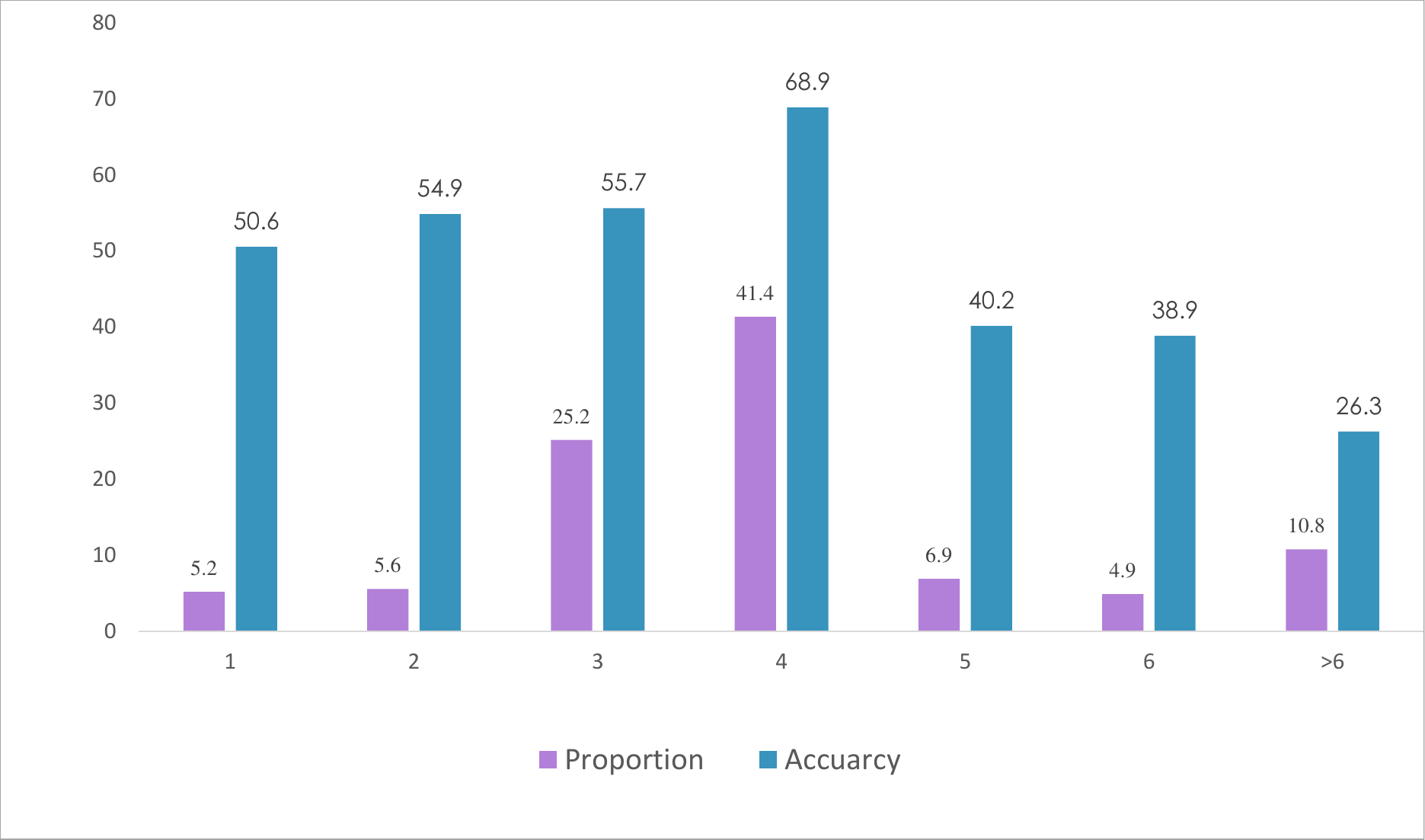}
    \end{center}
\vspace{-0.31cm}
\caption{Proportion and accuracy of samples with different formula sketch lengths in formula prediction task.}
\label{fig:sketch dist}
\end{figure}


\section{Error Analysis of Formula Prediction} \label{appendix:fp error analysis}
Figure~\ref{fig:sketch dist} presents the proportion and accuracy regarding different formula sketch lengths in prefix order~(parentheses excluded). As shown, sketch length $3$ and $4$ account for two-thirds of formulas, since length $3$ is typical for binary operations like \texttt{C4-B4}, and length $4$ is a common pattern for aggregation functions like \texttt{SUM(B4:C5)}. Thus, the accuracy of length $3$/$4$ is higher than shorter sketch length $1$/$2$ since more samples in its length are seen in training. And for longer formulas (\textgreater$6$), a significant performance drop occurs because complex nested references and calculations may be involved when the sketch gets longer.

To further analyze the errors in formula prediction, we randomly pick $100$ false generation results in dev set and divide these errors into three groups: (i) sketch failure~($54\%$): a wrong sketch is generated, which occurs more frequently when the formula gets longer and nested. A typical case is the formula with function \texttt{IF}, involving multiple arguments and nested calculations; (ii) reference unreachable~($27\%$): referenced cells are not in the sequence since we only consider the cells on the same row/column of the target cell as input; (iii) reference failure~($19\%$): wrong referenced cells are selected, which often occurs at the start or end of a cell range. Future works may improve formula prediction in these directions: handling long nested formulas, inputting more cells of table matrix as reference candidates conquering memory issues, and designing a module to match generated sketch with input table cells more accurately.

\section{Real examples of spreadsheet tables with formulas}
Here we show several real examples for spreadsheet tables in Figure [4-6].
\begin{figure*}[]
    \begin{center}
    \includegraphics[width=6.2in]{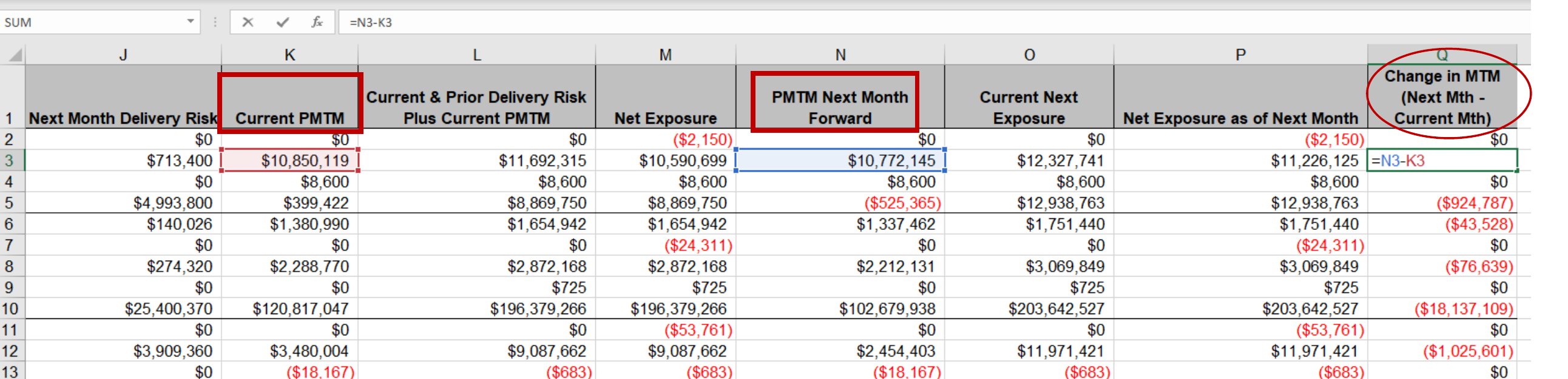}
    \end{center}
\caption{Example 1 with a substraction column.}
\vspace{-0.31cm}
\end{figure*}

\begin{figure*}[]
    \begin{center}
    \includegraphics[width=6in]{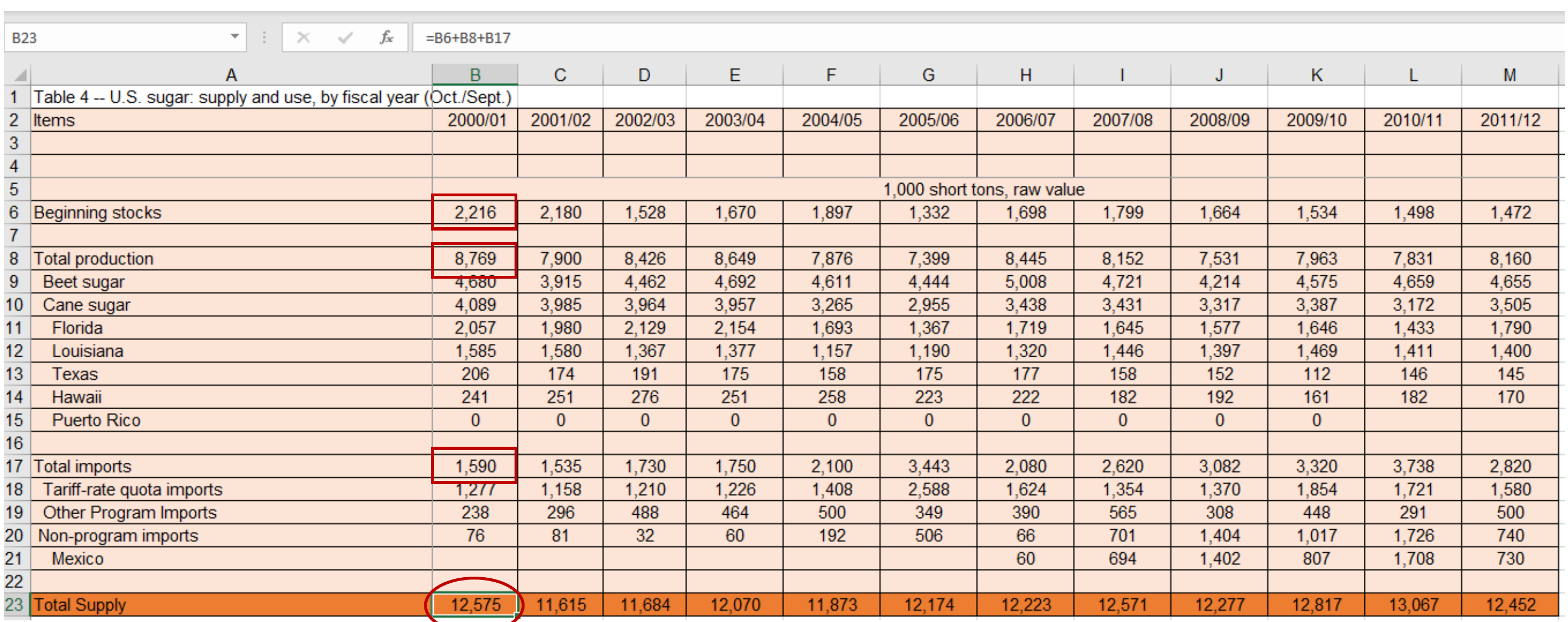}
    \end{center}
\caption{Example 2 with a total row.}
\vspace{-0.31cm}
\end{figure*}

\begin{figure*}[]
    \begin{center}
    \includegraphics[width=6in]{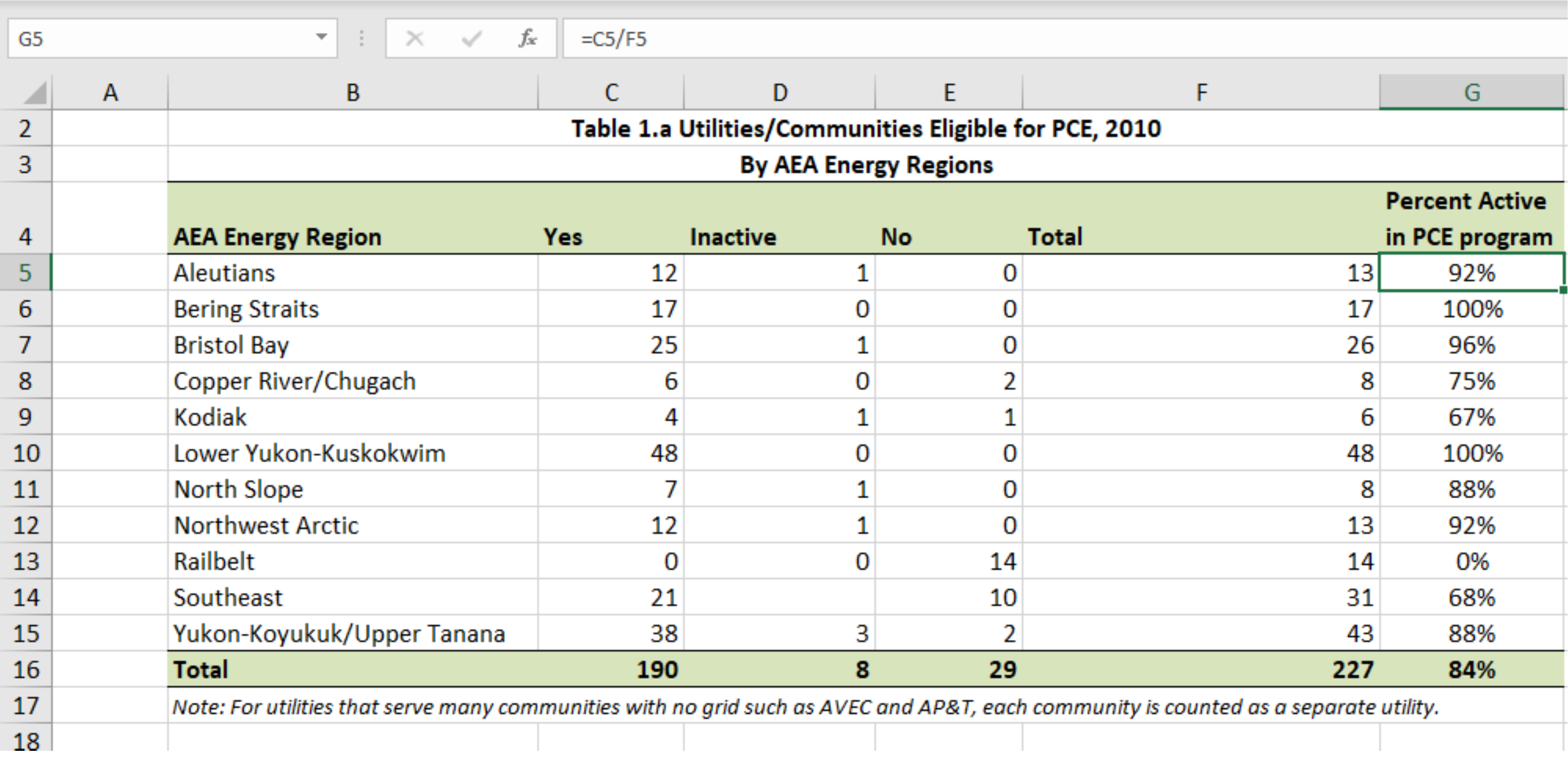}
    \end{center}
\caption{Example 3 with a total row and a proportion column.}
\vspace{-0.31cm}
\end{figure*}

\section{Real examples of formula prediction on Enron}
We also developed an Excel plug-in to run formula prediction powered by ForTaP. We simulate that ForTap suggests formulas for a user when she is editing a spreadsheet. Here we show several formula prediction demonstrations on Enron test set in Figure [7-11]. For the fist case, we tried different column names, and the results are promising and robust.

\begin{figure*}[]
    \begin{center}
    \includegraphics[width=6in]{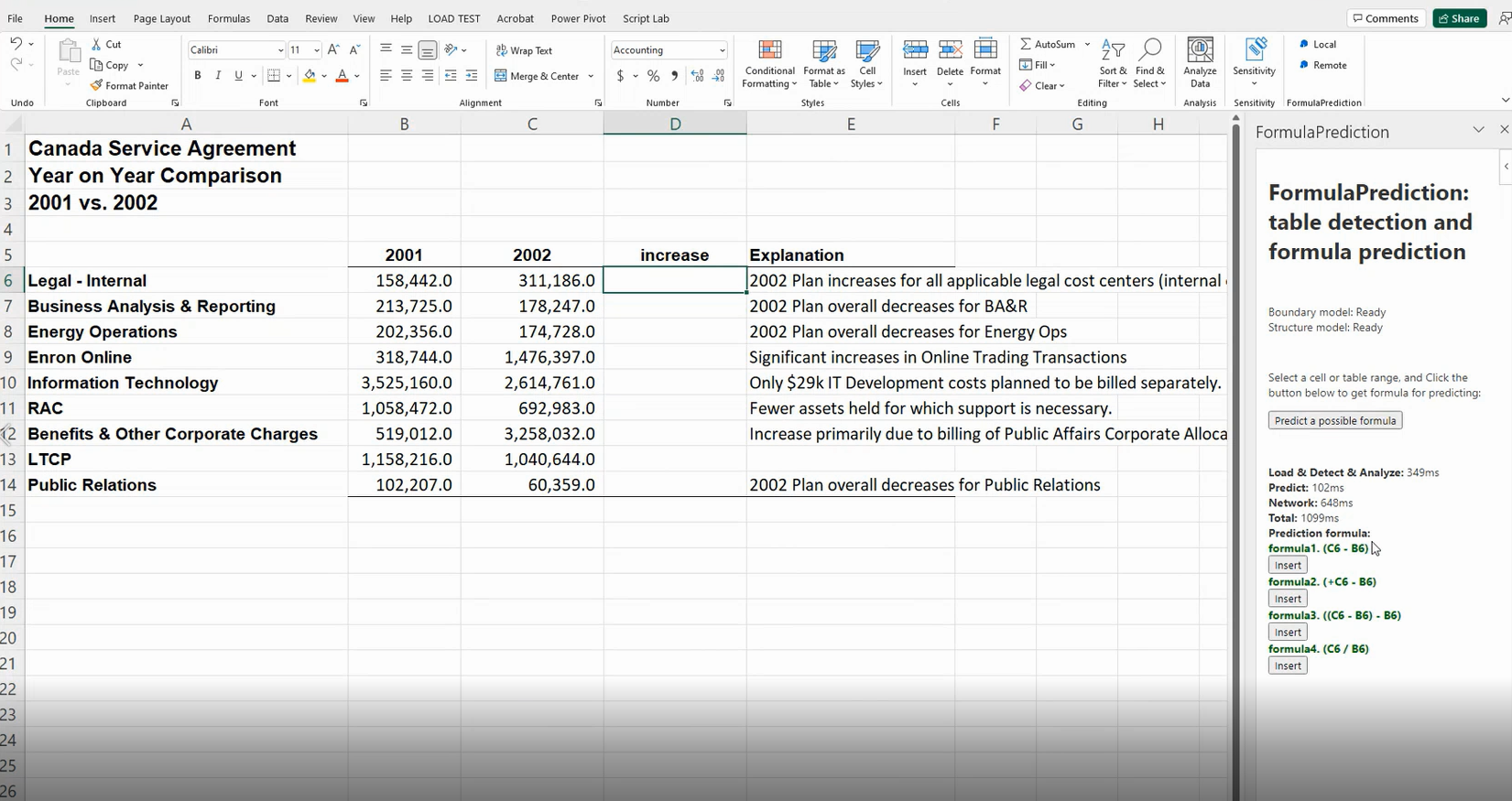}
    \end{center}
\caption{Example 1 modified on Enron test set for formula prediction.}

\end{figure*}

\begin{figure*}[]
    \begin{center}
    \includegraphics[width=6in]{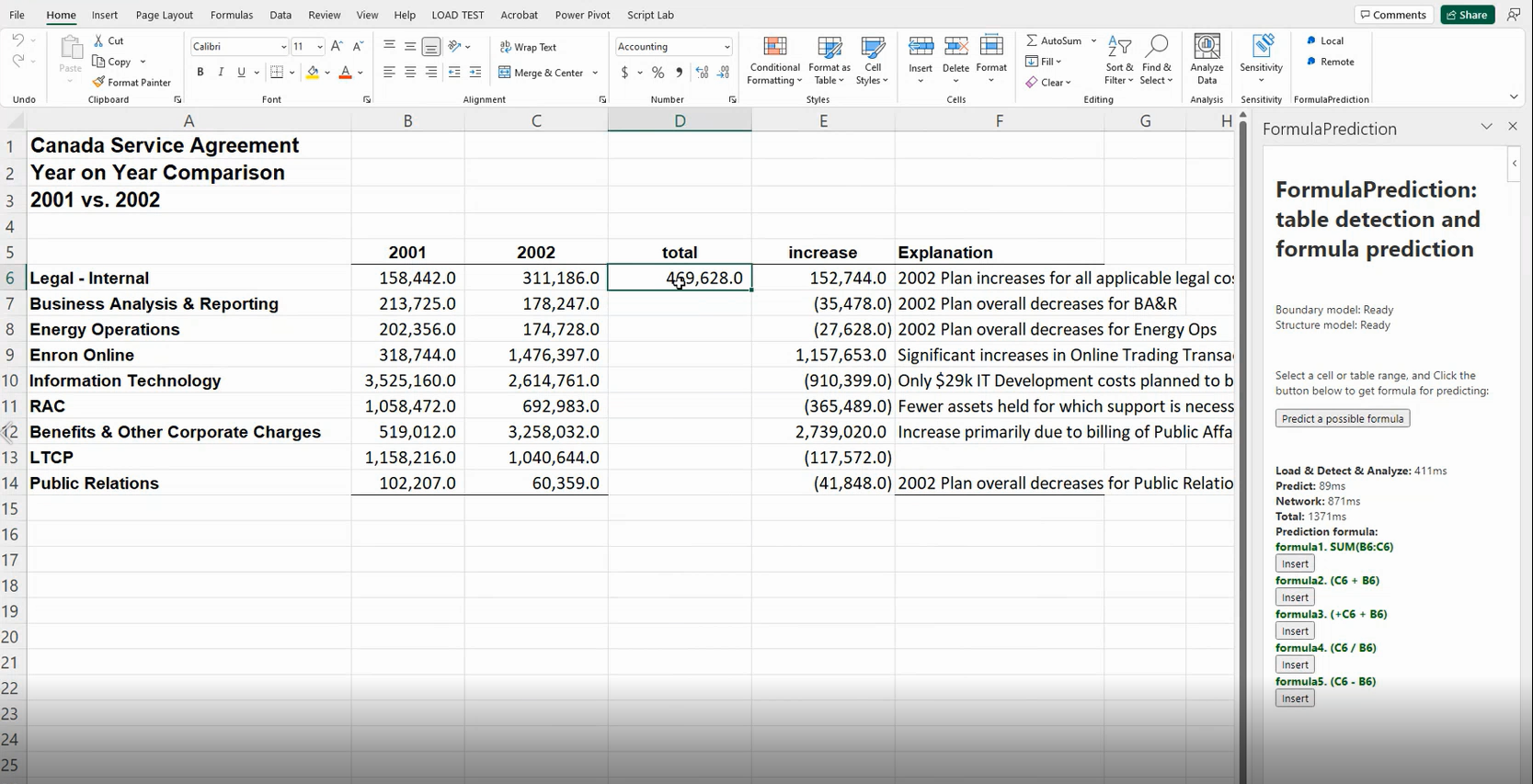}
    \end{center}
\caption{Example 2 modified on Enron test set for formula prediction.}

\end{figure*}

\begin{figure*}[]
    \begin{center}
    \includegraphics[width=6in]{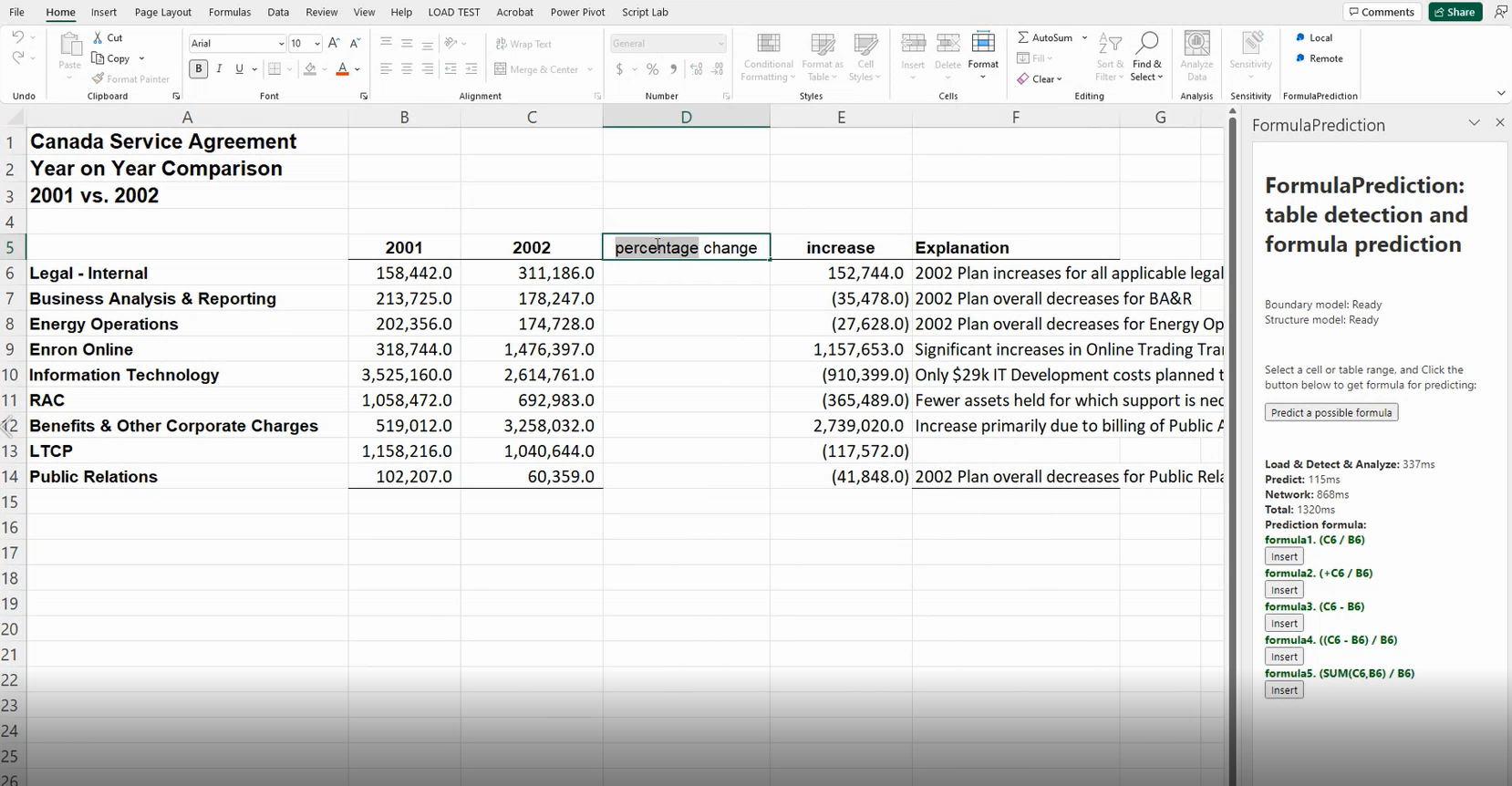}
    \end{center}
\caption{Example 3 modified on Enron test set for formula prediction.}

\end{figure*}

\begin{figure*}[]
    \begin{center}
    \includegraphics[width=6in]{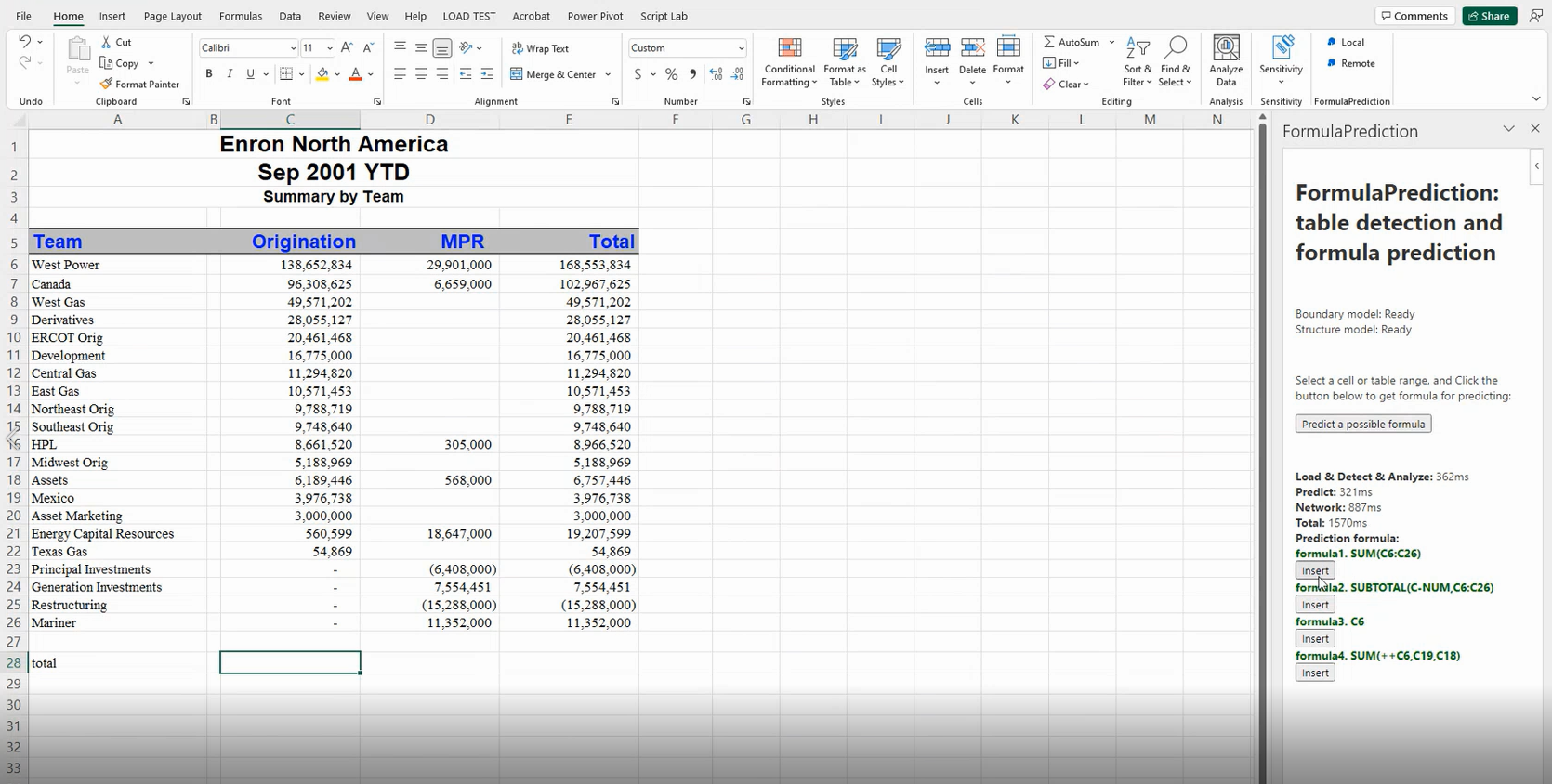}
    \end{center}
\caption{Example 4 modified on Enron test set for formula prediction.}

\end{figure*}

\begin{figure*}[]
    \begin{center}
    \includegraphics[width=6in]{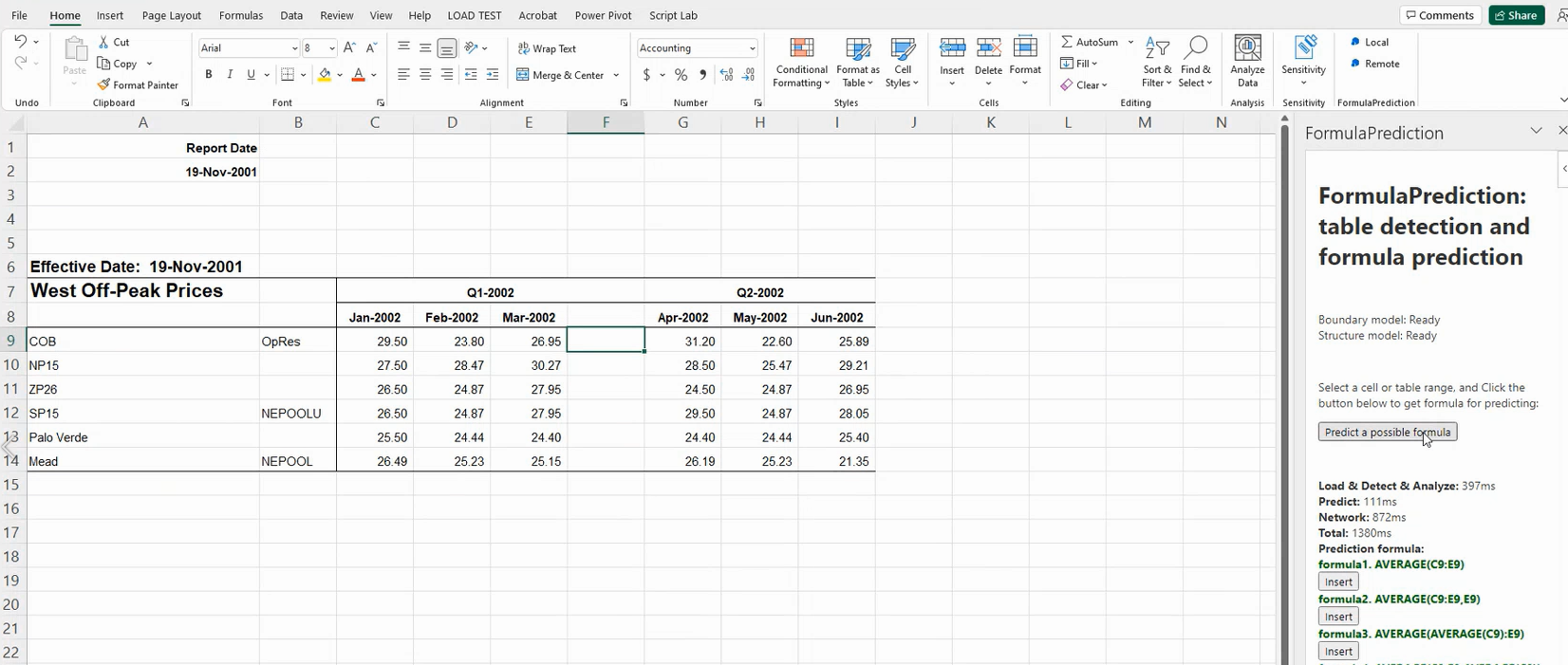}
    \end{center}

\caption{Example 5 on Enron test set  for formula prediction.}

\end{figure*}

\end{document}